\documentclass[aps,pra,twocolumn,groupedaddress,showpacs,
  amsmath,amssymb,amsfonts,floats,floatfix,raggedbottom,nobalancelastpage]{revtex4}
\usepackage{epsfig} 
\usepackage{psfrag}
\usepackage{color}
\definecolor{purple}{rgb}{1,0,1}

\begin{document}

\title{Quantum Monte Carlo study of the visibility of one-dimensional
  Bose-Fermi mixtures}  
\author{C. N. Varney$^1$, V. G. Rousseau$^2$, and R. T. Scalettar$^1$\\
$^1${\it Physics Department, University of California, Davis,
  California 95616, USA}\\ $^2$ {\it Instituut-Lorentz, Universiteit Leiden,
  Postbus 9504, 2300 RA Leiden, The Netherlands}\\
  (Received 9 November 2007)}

\begin{abstract}
The study of ultracold optically trapped atoms has opened new vistas
in the physics of correlated quantum systems.  Much attention has now
turned to mixtures of bosonic and fermionic atoms.  A central puzzle
is the disagreement between the experimental observation of a reduced
bosonic visibility ${\cal V}_b$, and quantum Monte Carlo (QMC)
calculations which show ${\cal V}_b$ increasing.  In this paper, we
present QMC simulations which evaluate the density profiles and ${\cal
V}_b$ of mixtures of bosons and fermions in one-dimensional optical
lattices.  We resolve the discrepancy between theory and experiment by
identifying parameter regimes where ${\cal V}_b$ is reduced, and where
it is increased.  We present a simple qualitative picture of the
different response to the fermion admixture in terms of the superfluid
and Mott-insulating domains before and after the fermions are
included.  Finally, we show that ${\cal V}_b$ exhibits kinks which are
tied to the domain evolution present in the pure case, and also
additional structure arising from the formation of boson-fermion
molecules, a prediction for future experiments.
\end{abstract}


\maketitle


It has been widely suggested that the strong correlations responsible
for magnetism, superconductivity, and the metal-insulator transition
in the solid state can be studied via ultracold optically trapped
atoms.  Indeed, this idea has been successfully realized in the
context of both bosonic and fermionic atoms.  In the former case, the
transition between condensed (superfluid) and insulating phases was
demonstrated through the evolution of the interference pattern after
the release and expansion of the gas \cite{Nature.415.39}.  Initial
studies focused on the height \cite{Nature.415.39} and width
\cite{PhysRevLett.92.130403} of the central interference peak, with
later work looking at the visibility ${\cal V}$, which measures the
difference between the maxima and minima of the momentum distribution
function $S({\bf k})$ \cite{PhysRevLett.95.050404, PhysRevA.72.053606,
PhysRevLett.98.180404}.  Interesting ``kinks'' are observed in ${\cal
V}$ which are associated with the redistribution of the density as the
superfluid shells evolve into insulating regions
\cite{PhysRevA.67.031602, PhysRevLett.95.220402}.  For trapped
fermions \cite{PhysRevA.68.011601,PhysRevLett.92.160601,
PhysRevLett.93.120407, PhysRevLett.94.080403,PhysRevlett.94.210401},
Mott phases could also form \cite{PhysRevLett.91.130403,
PhysRevA.69.053612, OptCommun.243.33}, however, without a signal in
$S({\bf k})$.  Instead, the evolution of the kinetic energy has been
proposed as a means to pinpoint the transition
\cite{PhysRevB.73.121103}.

Attention has turned at present to multicomponent systems, which offer
a rich set of phenomena including Bose-Einstein condensation (BEC)-BCS
crossover for two attractive fermionic species and
Fulde-Ferrell-Larkin-Ovchinnikov phases in situations with two
imbalanced fermion populations.  Two recent experimental papers report
the effect of the addition of fermionic $^{40}$K atoms on the
visibility of bosonic $^{87}$Rb in a three-dimensional trap
\cite{PhysRevLett.96.180402, PhysRevLett.96.180403}.  The basic result
is a decrease in the bosonic visibility ${\cal V}_b$ driven by the
fermion admixture.  A large number of qualitative explanations has
been put forth for this phenomenon, including the localization of the
bosons by the random fermionic impurities, the segmentation of the
bosonic superfluid, the adiabatic heating of the bosonic cloud when
the lattice depth is increased in the presence of the two species, an
enhanced bosonic mass due to the coupling to the fermions, and the
growth of Mott-insulating regions.  A fundamental difficulty is that
exact quantum Monte Carlo (QMC) calculations show an increase in
${\cal V}_b$ \cite{PhysRevA.77.023608}, in disagreement with
experiment.  In this paper we resolve this issue.

The behavior of Bose-Fermi mixtures has attracted considerable
theoretical attention.  The Hamiltonian was derived and its parameters
linked to experimental quantities by Albus {\em et al.}
\cite{PhysRevA.68.023606}. The equilibrium phase diagram has been
studied using mean-field theory and Gutzwiller decoupling
\cite{PhysRevA.68.023606,PhysRevLett.93.190405, arXiv:0708.3241},
perturbation theory \cite{PhysRevLett.93.190405}, dynamical mean-field
theory (DMFT) \cite{arXiv:0708.3241}, exact diagonalization
\cite{PhysRevA.69.021601}, quantum Monte Carlo methods
\cite{PhysRevLett.96.190402, PhysRevA.77.023608, PhysRevB.75.132507,
PhysRevA.76.043619}, and density matrix renormalization group (DMRG)
\cite{PhysRevA.77.023608, PhysRevA.77.023601}. The results of these
studies include the observation of Mott-insulating phases at ``double
half-filling'', anticorrelated winding of the two species of quantum
particles, molecule formation, and precise determination of the
exponents characterizing correlation function decay in the different
phases. The behavior of the visibility was addressed by Pollet {\em et
al.}  \cite{PhysRevA.77.023608}, who find interesting nonmonotonic
structures with fermion density. However, ${\cal V}_b$ is always
increased relative to the pure case \cite{footnote}.

In this paper we explore the visibility of Bose-Fermi mixtures in one
dimension using QMC simulations with the canonical worm algorithm
\cite{JETP.87.310,PhysRevLett.96.180603, PhysRevE.73.056703}.  While
previous QMC studies have reported a growth of ${\cal V}_b$, we show
that a significant reduction, such as seen experimentally, is also
possible without invoking temperature effects
\cite{PhysRevA.77.023608}.  The enhancement (reduction) of ${\cal
V}_b$ caused by the disruption (inducement) of the bosonic
Mott-insulator phase by the boson-fermion interactions.  ${\cal V}_b$
also exhibits kinks reminiscent of those in the pure boson case.
In the subsequent sections we write down the Hamiltonian and
observables and briefly discuss the QMC algorithm.  We then present
the evolution of ${\cal V}_b$ with fermion concentration, its
interpretation in terms of the bosonic density profiles, and evidence
for the formation of a molecular superfluid in the trap center.

\begin{figure}[t]
\psfrag{nb}{$n_b$} \psfrag{xi / a}{$x_i / a$}
\includegraphics[height=\columnwidth,angle=-90]{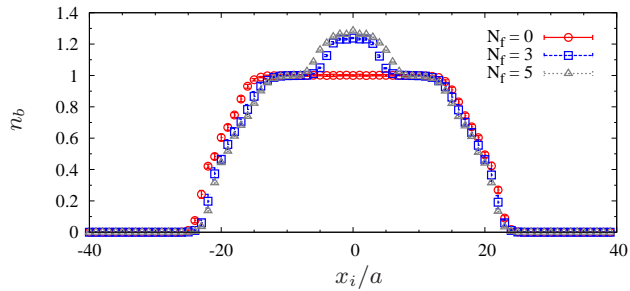}
\caption{(Color online) 
  Comparison of the density profiles at $U_{bb}=8.3t$ and
  $U_{bf}=-5.0t$ for $N_f=0, 3, 5$ fermions on an 80-site chain with
  40 bosons.  The Mott insulator at the trap center for the pure
  bosonic case is destroyed by the addition of fermions.  This drives
  the increase in the visibility. 
\label{fig:rho_comp_Ubb8.2}
}
\end{figure}

The one-dimensional (1D) Bose-Fermi Hubbard
Hamiltonian is \cite{PhysRevA.68.023606}
\begin{align}
  \label{hamiltonian}
  \begin{aligned}
  H =&-t_b \sum_j ( \hat{b}_j^\dagger \hat{b}_{j + 1}^{\phantom
    \dagger} + {\rm h.c.} ) 
  -t_f \sum_j ( \hat{c}_j^\dagger \hat{c}_{j + 1}^{\phantom \dagger}
  + {\rm h.c.} ) \\ 
  &+ W \sum_i x_i^2 (\hat{n}_b^{(i)} + \hat{n}_f^{(i)})\\ 
  &+ \frac{U_{bb}}{2} \sum_i \hat{n}_b^{(i)} ( \hat{n}_b^{(i)} - 1 ) +
  U_{bf} \sum_i \hat{n}_b^{(i)} \hat{n}_f^{(i)} ,
  \end{aligned}
\end{align}
where $\hat{b}^{\phantom \dagger}_j$ ($\hat{b}^\dagger_j$) and
$\hat{c}^{\phantom \dagger}_j$ ($\hat{c}^\dagger_j$) are the
annihilation (creation) operators of the bosons and (spin-polarized)
fermions at lattice site $j$, respectively, and $\hat{n}_b^{(i)} =
\hat{b}_i^\dagger \hat{b}_i^{\phantom \dagger}$, $\hat{n}_f^{(i)} =
\hat{c}_i^\dagger \hat{c}_i^{\phantom \dagger}$ are the corresponding
number operators. The first two terms of Eq.~\eqref{hamiltonian}
describe bosonic and fermionic nearest-neighbor hopping. The curvature
of the trap is $W$, and the coordinate of the $j$th site is given by
$x_j = j a$, where $a$ is the lattice constant. $U_{bb}$ and $U_{bf}$
are the on-site boson-boson and boson-fermion interactions.  In this
work we consider 80-site chains with the nearest-neighbor hopping set
to be identical for fermions and bosons $(t_b = t_f = t = 1)$ and
trapping potential $W = 0.01t$.

In the canonical worm algorithm \cite{JETP.87.310,
PhysRevLett.96.180603, PhysRevE.73.056703} employed in our
calculation, operator expectation values are sampled through the
introduction of open-ended world lines that extend over equal
imaginary time into a path integral expression for the partition
function.  The properties we study include the kinetic, potential, and
trap energies, the density profiles, and the visibility,
\begin{align}
{\cal V} &= \frac{S_{\rm max}-S_{\rm min}} {S_{\rm max}+S_{\rm min}}\,\,
,
\label{vis}
\end{align}
where $S_{\rm max}$ and $S_{\rm min}$ are the maximum and minimum
values of momentum distribution function,
\begin{align}
S(k) &= \dfrac{1}{L}\sum_{j,\,l} {\rm e}^{i k (x_j - x_l)}\langle
    \hat{b}_j^{\dagger} \hat{b}_l^{\phantom \dagger}\rangle.
\end{align}


\begin{figure}[t]
\psfrag{v}{${\cal V}$}
\psfrag{Nb}{$N_b$}
\psfrag{nb}{$n_b$}
\psfrag{xi / a}{$x_i / a$}
\psfrag{Sbmax}{$S_{\rm max}$}
\psfrag{(a)}{\footnotesize{(a)}}
\psfrag{(b)}{\footnotesize{(b)}}
\includegraphics[width=\columnwidth]{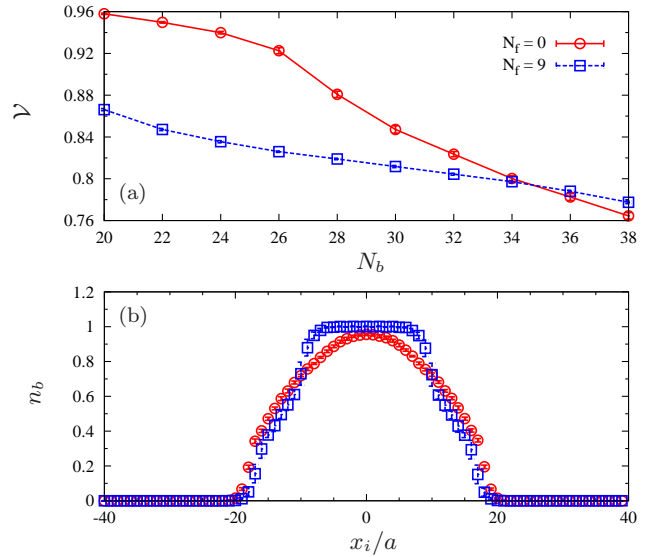}
\caption{(Color online) 
  (a) Bosonic visibility ${\cal V}_b$ as a function of the number of
  bosons for $N_f = 0$ and 9 fermions with fixed $U_{bb}=12.0t$ and
  $U_{bf}=-5.0t$. (b) Bosonic density profiles for $N_b = 26$ bosons
  and $N_f = 0, 9$ fermions. The addition of the fermions induces
  Mott-insulating behavior in the bosons.  The key consequence is a
  decrease in ${\cal V}_b$ for $N_f=9$ relative to $N_f=0$, similar to
  that seen in the experiments.
\label{fig:visibility_nbcomp}
} 
\end{figure}

The enhanced visibilities with fermion concentration reported
previously \cite{PhysRevA.77.023608} are in contrast with the trend to
reduced ${\cal V}_b$ measured experimentally
\cite{PhysRevLett.96.180402, PhysRevLett.96.180403}.  In
Fig.~\ref{fig:rho_comp_Ubb8.2}, we see the origin of this effect in a
system with 40 bosons: the visibility enhancement at large $U_{bb}$ is
caused by the destruction of the Mott phase at the trap center by the
fermions.  It is natural to conjecture that if $n_b^{(i)}<1$ at the
trap center the additional attraction due to the fermions could induce
Mott-insulating behavior and reduce ${\cal V}$.  In
Fig.~\ref{fig:visibility_nbcomp}(a), we show that this expectation is
correct.  Here, we fix $U_{bb}=12t$ and increase $N_b$ for both the
pure case and for a system with fermion number fixed at $N_f = 9$ and
boson-fermion interaction at $U_{bf} = -5t$.
What we observe is that in a window where the boson central density is
approaching $n_b^{(i)}=1$ the bosonic visibility is decreased by the
presence of the fermions.  The cause is clear: if the bosons are
poised just below Mott-insulating behavior, then the fermions can induce it.
This is supported by a comparison of the density profiles in
Fig.~\ref{fig:visibility_nbcomp}(b).

The primary mechanism through which fermions affect ${\cal V}_b$ is
the local adjustment of the site energy and hence of the local bosonic
density.  This is an effect which occurs regardless of the
dimensionality.  Hence, we expect aspects of our conclusions to be
relevant to experiments in higher dimension
\cite{PhysRevA.77.023608}.
While we have shown a decreased visibility similar to that seen
experimentally, the enhancement of visibility may be the more generic
behavior in one dimension.  In the one-dimensional ``state diagram''
of the purely bosonic case \cite{PhysRevLett.89.117203} the area of
parameter space occupied by the phase with a Mott plateau of
$n_b^{(i)}=2$ is very narrow.  Thus the prospect for the fermions to
drive the system into this phase is limited.

\begin{figure}[t]
\psfrag{Ubb / t}{$U_{bb}/t$}
\psfrag{xi / a}{$x_i/a$}
\psfrag{Smax}{$S^b_{\rm max}$}
\psfrag{nb}{$n_b$}
\psfrag{v}{${\cal V}$}
\psfrag{(a)}{\footnotesize{(a)}}
\psfrag{(b)}{\footnotesize{(b)}}
\includegraphics[width=\columnwidth]{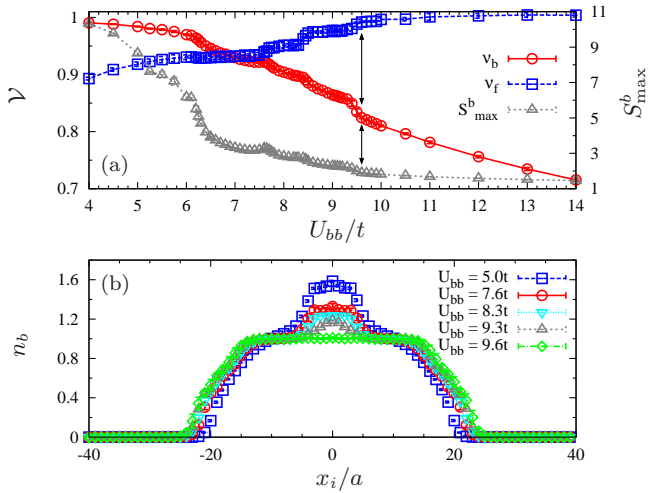}
\caption{(Color online)  
  (a) Bosonic and fermionic visibilities and bosonic $S_{\rm max}$ as
  functions of $U_{bb}/t$ for a system with 40 bosons, 3 fermions, and
  $U_{bf}=-5.0t$.  The ``plateau'' regions where the rate of reduction
  of ${\cal V}$ is reduced are due to freezing of the density profiles
  (see text). The fast decrease after $U_{bb}/t\approx9.3$ is due to
  the formation of a Mott-insulating region in the central core, which
  is fully formed and indicated by the arrow at $U_{bb}/t = 9.6$.  (b)
  Boson density profiles at five different values of $U_{bb}/t$.
\label{fig:visnf3}
}
\end{figure}

\begin{figure}[b]
\psfrag{Ubb / t}{$U_{bb}/t$}
\psfrag{Etrap}{$E_{\rm trap}$}
\psfrag{Eint}{$E_{\rm int}$}
\psfrag{Ebkin}{$E_{\rm kin}^b$}
\psfrag{Efkin}{$E_{\rm kin}^f$}
\psfrag{(a)}{\footnotesize{(a)}}
\psfrag{(b)}{\footnotesize{(b)}}
\includegraphics[width=\columnwidth]{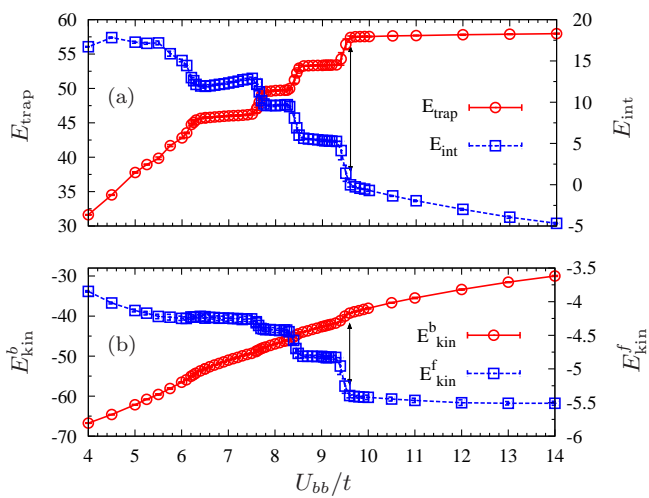}
\caption{(Color online) 
  (a) Trapping ($E_{\rm trap}$) and interaction ($E_{\rm int}$) energies
  as functions of $U_{bb}/t$ for the system of Fig.~\ref{fig:visnf3}.
  (b) Bosonic ($E_{\rm kin}^b$) and fermionic ($E_{\rm kin}^f$)
  kinetic energies.
\label{fig:energynf3}
}
\end{figure}

In the case of a pure bosonic system \cite{PhysRevLett.95.220402}, the
change in visibility with the boson-boson interaction strength
$U_{bb}$ is not smooth, but is accompanied by ``kinks.'' These kinks are
associated with a freezing of the density profile when the transfer of
the bosonic density from the trap center is interrupted by the
formation of Mott insulator shoulders.  In Fig.~\ref{fig:visnf3}, the
behavior of the visibilities and density profiles with $U_{bb}$ in the
presence of fermions is shown.  ${\cal V}_b$ decreases with $U_{bb}$
as the interactions reduce the quasicondensate fraction $S^{b}_{\rm
max}$.  Conversely, the interactions enhance $S^f_{\rm max}$ and
${\cal V}_f$ increases with $U_{bb}$.  There are, however, additional
kinks in the case when fermions are present whose origin we shall
discuss below.  Figure~\ref{fig:energynf3} helps to quantify this
freezing by showing the evolution of the trap, interaction, and
kinetic energies with $U_{bb}$.  These energies exhibit a sequence of
plateaus and rapid drops corresponding to the kink locations in
Fig.~\ref{fig:visnf3}.

\begin{figure}[t]
\psfrag{Ubb / t}{$U_{bb}/t$}
\psfrag{Sbmax}{$S^b_{\rm max}$}
\psfrag{nb}{$n_b$}
\psfrag{Etrap}{$E_{\rm trap}$}
\psfrag{Eint}{$E_{\rm int}$}
\psfrag{vb}{${\cal V}_b$}
\psfrag{(a)}{\footnotesize{(a)}}
\psfrag{(b)}{\footnotesize{(b)}}
\psfrag{(c)}{\footnotesize{(c)}}
\includegraphics[width=\columnwidth]{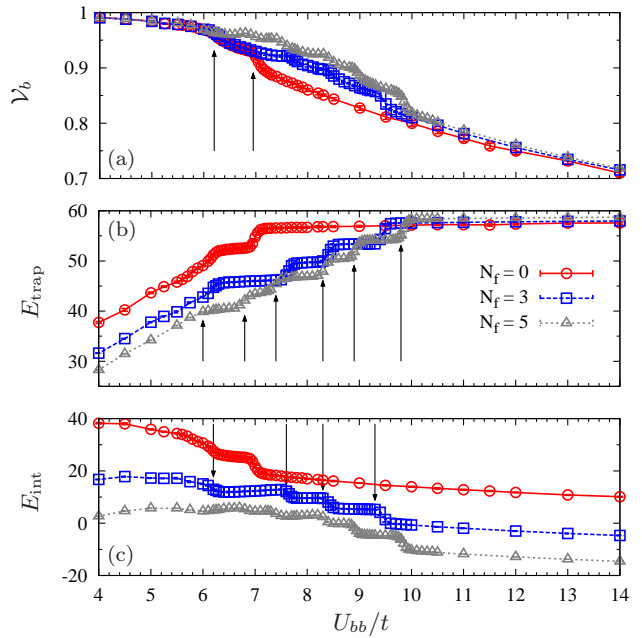}
\caption{(Color online)  
  Comparison of (a) bosonic visibility ${\cal V}_b$, (b) the trapping
  energy, and (c) the interaction energy for $N_f = 0$, $3$, and $5$
  fermions on an 80-site chain with 40 bosons and fixed $U_{bf} =
  -5.0t$. The arrows in panels (a), (c), and (b), respectively, denote
  the locations of the kinks (the onset of rapid change in energy and
  visibility) for $N_f = 0$, $3$, and $5$ fermions.
\label{fig:visenergycomp}
}
\end{figure}

Figure~\ref{fig:visenergycomp}(a) compares the visibility evolution
for the pure bosonic case ($N_f=0$) with two different fermion numbers
$N_f=3$ and $5$.  For $N_f = 3$, the kink at lowest $U_{bb}=6.1t$ is
associated with the initial emergence of the Mott shoulders. This kink
coincides with one in the pure bosonic case $N_f=0$ because the
shoulders form outside the regions occupied by the fermions at the
trap center. For $N_f =5$, the width of the fermion density is
comparable to the size of the bosonic superfluid in the center of the
trap, and the kink at $U_{bb}=6.0t$ signifies a freezing of the
bosonic density but not the formation of the Mott shoulders.  Instead,
the kink visible at $U_{bb} \approx 6.8t$ is responsible for the
formation of the Mott shoulders.  This shift to higher $U_{bb}$ is
expected since the attractive $U_{bf}$ delays the transfer of bosonic
density out of the center.  Figures~\ref{fig:visenergycomp}(b) and
\ref{fig:visenergycomp}(c) compare the components of the energy. Each
plateau signifies that the bosonic and fermionic densities are frozen
over the range in $U_{bb}$.  Past experiments
\cite{PhysRevLett.96.180402,PhysRevLett.96.180403} did not have the
resolution to exhibit the kinks we have seen in these simulations;
however, they might be seen with improved accuracy.

\begin{figure}[t]
\psfrag{r}{$n$}
\psfrag{xi / a}{$x_i / a$}
\psfrag{(a)}{\footnotesize{(a)}}
\psfrag{(b)}{\footnotesize{(b)}}
\psfrag{(c)}{\footnotesize{(c)}}
\psfrag{(d)}{\footnotesize{(d)}}
\includegraphics[height=\columnwidth,angle=-90]{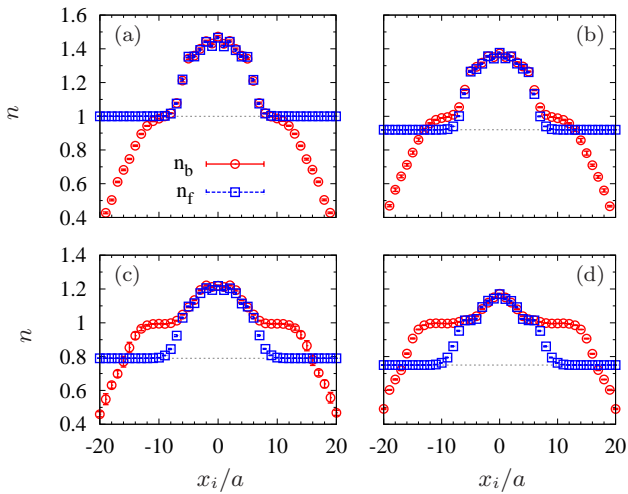}
\caption{(Color online)  
  Bosonic and fermionic density profiles for (a) $U_{bb}=6.3t$, (b)
  $U_{bb}=7.2t$, (c) $U_{bb}=8.6t$, and (d) $U_{bb}=9.5t$ with 5
  fermions, 40 bosons, and fixed $U_{bf} = -5.0t$. The fermionic
  density is offset and the dashed gray line indicates zero
  density. The densities match in the center of the
  trap, with the region of coincidence decreasing as $U_{bb}$
  increases. 
  \label{fig:rhocomp}
}
\end{figure}

We also note in Fig.~\ref{fig:visenergycomp} that the number of
plateaus is directly related to the number of fermions in the
Bose-Fermi mixture and that each plateau is roughly the same size,
indicating that bound pairs of bosons and fermions are being destroyed
as $U_{bb}$ is increased. This conclusion is substantiated in
Fig.~\ref{fig:rhocomp}, where we compare the bosonic density with the
fermionic density. The fermionic density is offset by a constant to
emphasize the near perfect overlap in the densities near the center of
the trap, indicating that the trap center is populated by a molecular
superfluid (MSF).  Indeed, at the weakest coupling,
Fig.~\ref{fig:rhocomp}(a), the fermion density precisely equals the
excess boson density above the commensurate Mott value $n_b^{(i)}=1$.
When $U_{bb}$ is increased, moving from one plateau to another in
Fig.~\ref{fig:visenergycomp}, the MSF region in the center of the trap
shrinks. For the kink at highest $U_{bb}$ ($\approx 9.8t$ for
$N_f=5$), the MSF region is destroyed, the bosonic density is a Mott
insulator, and the fermionic visibility ${\cal V}_f \to 1$.



In summary, we have shown that the visibility of Bose-Fermi mixtures
can be enhanced or reduced by the boson-fermion interactions depending
on whether the bosonic density in the pure case is above or below
commensuration.  This result resolves a fundamental disagreement
between experiment and QMC simulations.  There are numerous kinks in
the visibility and the different energies that result from freezing of
the density profiles.  While our bosonic component is sufficiently
large so that our results are converged with respect to lattice size,
the number of fermions is much smaller.  It is possible that the kinks
will merge together and be less easy to observe in a larger system.
The density profiles near the kinks show direct evidence for a
molecular superfluid in the center of the trap and that a larger
$U_{bb}$ is required to destroy the bound pairs with larger $N_f$ and
subsequently induce Mott-insulating behavior.

Supported under ARO Award W911NF0710576 with funds from the DARPA OLE
Program, NSF PHY05-51164 (preprint NSF-KITP-07-195), with useful input 
from T. Mavericks.

\bibliography{visibility,bfexpt,bftheory,optical,misc}

\begin{thebibliography}{10}
\bibitem{footnote}
For one set of parameters, ${\cal V}$ for non-zero $N_f$ falls below the
  $N_f=0$ value, but only when $N_f$ exceed $N_b$. In addition, the decrease is
  less than ten percent.
\end{thebibliography}

\end{document}